\documentclass[journal=jacsat,manuscript=article]{achemso}
\usepackage[version=3]{mhchem} 

\usepackage{xcolor}

\author{Indrajit Maity}
\email{indrajit.maity@newcastle.ac.uk}
\affiliation{Department of Chemistry, School of Natural and Environmental Sciences, Newcastle University, Newcastle upon Tyne, NE1 7RU, UK.}
\author{Johannes Lischner}
\affiliation{Departments of Materials and Physics and the Thomas Young Centre for Theory and Simulation of Materials, Imperial College London, South Kensington Campus, London, SW7 2AZ, UK.}
\author{Arash A. Mostofi}
\affiliation{Departments of Materials and Physics and the Thomas Young Centre for Theory and Simulation of Materials, Imperial College London, South Kensington Campus, London, SW7 2AZ, UK.}
\author{\'{A}ngel Rubio}
\email{angel.rubio@mpsd.mpg.de}
\affiliation
{Initiative for Computational Catalysis and Center for Computational Quantum Physics, Flatiron Institute, Simons Foundation, New York City, New York 10010, USA.}
\altaffiliation{Max Planck Institute for the Structure and Dynamics of Matter, Luruper Chaussee 149, 22761 Hamburg, Germany.}

\title[An \textsf{achemso} demo]
  {Origin of trapped intralayer Wannier and charge-transfer excitons in moiré materials}

\begin{document}







\begin{abstract}
Moiré materials provide a versatile platform for engineering excitons, enabling next-generation optoelectronic applications. Continuum models are widely used to study moiré excitons due to their efficiency, but they often disagree with ab initio many-body approaches, as seen for intralayer excitons in WS$_2$/WSe$_2$ heterobilayers. Here, we resolve these discrepancies using an atomistic, quantum-mechanical framework based on the Bethe-Salpeter equation with Wannier functions as the electronic structure basis, showing that dielectric screening from hBN encapsulation is essential to reproduce experimentally observed exciton features. Our analysis reveals that exciton behavior emerges from a subtle competition between Wannier and charge-transfer characters, driven by stacking-dependent intralayer bandgap variations and environment-tuned electron–hole interactions. We show that the lowest-energy bright intralayer excitons are Wannier-like in WS$_2$/WSe$_2$ heterobilayers but charge-transfer-like in twisted WSe$_2$ homobilayers, despite comparable moiré sizes. These results establish atomistic modeling as a powerful tool for understanding and controlling excitonic phenomena in moiré materials.
\end{abstract}

\section{Introduction}
An exciton is a neutral quasiparticle that forms in a semiconductor or
insulator when an electron is excited from the valence band to the conduction
band, leaving behind a hole. The electron and hole are bound together by the
Coulomb interaction~\cite{Rohlfingelectron2000,Oniaelectronic2002}. Excitons
dominate the optical response of two-dimensional (2D) layered semiconductors,
such as semiconducting transition metal dichalcogenides
(TMDs)~\cite{Wangexcitons2018}. Moir\'{e} patterns created by stacking and
twisting 2D semiconducting TMDs offer unprecedented opportunities to manipulate
excitons. For example, moiré patterns can trap excitons in quantum-dot-like
states, enabling applications in optoelectronics such as
lasers~\cite{Chenjianglasing2024}, modulators and single-photon
emitters~\cite{Baekhighly2020}, and allowing the creation of artificial
lattices of interacting quasiparticles for quantum information
processing~\cite{Wangquantum2024,Honglishedding2020} and many-body
physics~\cite{Hongyimoire2017,Kennesmoire2021,Seylersignatures2019,Maurospin2020,
Baekhighly2020,Herreramoire2025,Kumlinsuper2025}. 

\begin{figure*}[htbp!]
    \centering
    \includegraphics[scale=0.5]{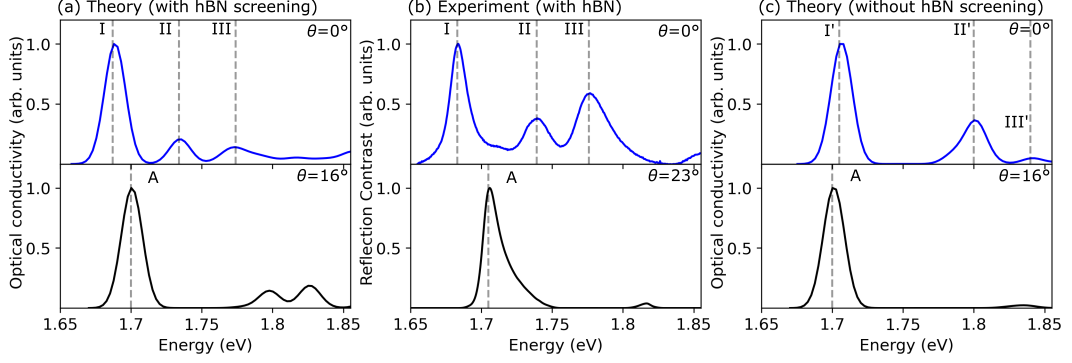}
    \caption{\textbf{Comparison of theoretical and
experimental~\cite{Jinobservation2019} results for intralayer excitons of
WSe$_{2}$ in WS$_{2}$/WSe$_{2}$}. At large twist angles, the exciton spectrum
is dominated by the A peak, similar to that of monolayer WSe$_{2}$. At small
twist angles, the A peak splits into three peaks, labeled I, II, and III,
separated by approximately 90 meV. (a) and (b) show theory and experiments with
hexagonal boron nitride (hBN) as the substrate, while (c) shows theory without
hBN.} \label{theoryvsexpt}
\end{figure*}

The experimental hallmark of \textit{intralayer moir\'{e} excitons} in
WS$_2$/WSe$_2$, a prototypical heterobilayer, is the appearance of three
absorption peaks around 1.7~eV in nearly aligned
samples~\cite{Jinobservation2019}. These peaks are separated by about 90~meV,
with the first peak redshifted by 20-30~meV compared to its position in
large-twist-angle WS$_2$/WSe$_2$. While the experimental signatures of
moir\'{e} excitons are firmly
established~\cite{Jinobservation2019,Sandhyahyperspectral2022,Kaidistinct2025,Zhaodynamic2021,Liuexcitonic2021},
the spatial distributions of the associated exciton wavefunctions remain highly
debated. This debate arises partly from the difficulty of imaging exciton confinement with meV energy and $\mathrm{\AA}$-scale spatial resolution~\cite{Sandhyahyperspectral2022,Sankarseeing2025,Dandumapping2024}, and partly from disagreements among theoretical studies on the nature of exciton wavefunctions, even though all reproduce the three moiré exciton peaks. Early continuum models used an effective
\textit{moiré potential} to modulate the center-of-mass motion of excitons in monolayer
WSe$_{2}$ to capture intralayer moir\'{e}
excitons~\cite{Wutopological2017,Jinobservation2019,Bremtunable2020}. By
adjusting the potential depth, the theory can reproduce the three moir\'{e}
exciton peaks seen in experiments. It also predicts that all three moir\'{e}
excitons are of the \textit{Wannier type}. In contrast, Naik and co-workers performed
large-scale ab initio GW plus Bethe–Salpeter Equation (GW-BSE) calculations and
found that the three excitons are not all of Wannier type. One of the three peaks has strong \textit{charge-transfer
character}~\cite{Naikintralayer2022}. These large-scale atomistic calculations
were made possible by approximating moir\'{e} electronic wavefunctions as a
coherent superposition of pristine unit cell wavefunctions, inspired by band
unfolding~\cite{Naikintralayer2022}. Some refined continuum models developed
subsequently did not find the charge-transfer
character~\cite{Changcontinuum2023}. Others that did still predicted exciton
localizations that disagreed with the results of Naik and
co-workers~\cite{Wangtwist2025}. For example, GW-BSE predicts the lowest bright
exciton to localize in the AA region (W above W and Se above
S)~\cite{Naikintralayer2022}, whereas improved continuum calculations find it
in the $\mathrm{B^{Se/W}}$ region (Bernal stacking with Se above
W)~\cite{Wangtwist2025}.  Nevertheless, both GW-BSE~\cite{Naikintralayer2022}
and continuum models~\cite{Wangtwist2025} predict the energy separation of the
three moir\'{e} exciton peaks to be about 180-200~meV, compared to the
experimentally observed value of ~93~meV. Additionally, the redshift of the
lowest bright moir\'{e} exciton energy remains unexplained. Despite extensive efforts, no existing theory has achieved a unifying and quantitatively accurate explanation of all experimentally observed moiré exciton features and their microscopic origin.

In our previous work~\cite{Maityatomistic2025}, we identified three intralayer moiré excitons in WS$_2$/WSe$_2$ for twist angles below 2$^\circ$. However, the bright excitons were not fully localized, unlike those reported in GW-BSE and continuum studies~\cite{Naikintralayer2022,Jinobservation2019,Wangtwist2025,Changcontinuum2023}, and one exhibited very weak oscillator strength. Because those calculations were limited to moiré lattice constants up to 7 nm~\cite{Maityatomistic2025}, smaller than the ~8.2 nm systems typically examined in experiments and previous theories. Addressing this gap is essential to a unified understanding of moiré excitons and their experimental signatures.

Motivated by these unresolved questions, we investigate the nature of low-energy moiré
\textit{intralayer excitons} in WSe$_2$ in WS$_2$/WSe$_2$ heterostructure by solving BSE based on atomistic simulations of experimentally
relevant moir\'{e} system sizes. We improve recently developed algorithms for computing moiré excitons with Wannier functions~\cite{Maityatomistic2025} by introducing an accurate interpolation scheme, allowing calculations on larger systems than previously possible. We demonstrate for the first time that dielectric screening from hBN encapsulation is crucial for reproducing the experimentally observed behavior of moiré intralayer excitons. In particular, the lowest bright exciton becomes completely localized at AA stacking region
and is of the Wannier type, while the highest bright exciton exhibits both
strong charge-transfer character and spatial delocalization when hBN screening is included. Without hBN screening, the charge-transfer
character of the highest bright exciton disappears entirely, and the
localization of the lowest bright exciton is significantly weaker. Our analysis reveals a new mechanistic insight: a subtle competition between local stacking-dependent \textit{intralayer band gaps} and the electron-hole Coulomb interaction that binds the exciton. Using an \textit{adiabatic
switching} approach, we show how atomic rearrangements in the moiré pattern and
hBN screening influence this competition. We show that stacking different materials can tune the ordering of Wannier and charge-transfer characters. In parallel-aligned WS$_2$/WSe$_2$ heterobilayers, the lowest bright intralayer exciton remains Wannier-like regardless of hBN screening. In contrast, in a 57.7$^\circ$ twisted WSe$_2$ bilayer, the lowest bright exciton is charge-transfer-like with hBN screening and Wannier-like without it, even though both systems have nearly identical moiré lattice constants.

All moiré structures were generated using the TWISTER package~\cite{Naiktwister2022}. Atomic relaxations were obtained using classical interatomic potentials within the LAMMPS package~\cite{lammps, Thompsonlammps2022, Zhouhandbook2017, Naikkolmogorov2019, Bitzekstructural2006}. Electronic structure calculations were carried out using the SIESTA package~\cite{Solersiesta2002,Troullierefficient1991, Perdewself1981}. Wannier functions were generated using the WANNIER90 code~\cite{Marzarimaximally2012, Pizziwannier902020,Souzamaximally2001}, and all exciton calculations were performed with a new version of the PyMEX package~\cite{Maityatomistic2025,Auckenthalerparallel2011, MarekElpa2014}. Screened Coulomb interactions were treated using the Rytova-Keldysh potential~\cite{Cudazzodielectric2011,Keldyshcoulomb1979,Berkelbachtheory2013,Chernikovexciton2014,Rytovascreened2018,Vancoulomb2018,Gorycarevealing2019}. See Supplementary Information (SI), Sec. A, for more technical details.

\section{Results and discussion}
\subsection{Intralayer excitons of WSe$_{2}$ in WS$_{2}$/WSe$_{2}$ heterobilayers}
\begin{figure*}[ht!]
    \centering
    \includegraphics[scale=0.45]{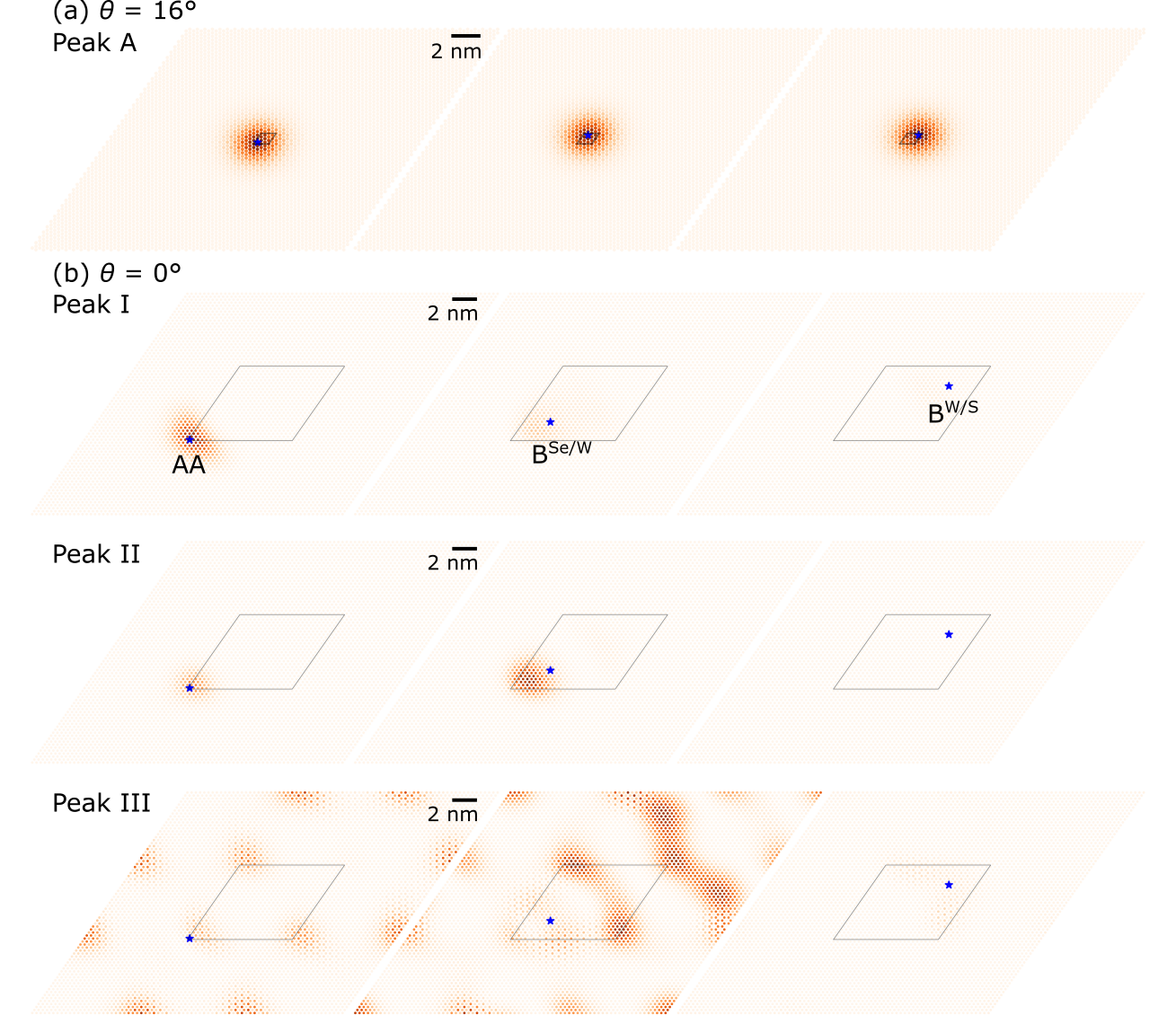}
    \caption{\textbf{Spatial distribution of intralayer exciton wavefunctions for a fixed hole in twisted WS$_{2}$/WSe$_{2}$ (with hBN screening)} (a) At $\theta = 16^\circ$, peak A exhibits the same behavior regardless of hole position.
    (b) At $\theta = 0^\circ$, peaks I, II, and III (as shown in Fig.~\ref{theoryvsexpt}(a)) exhibit distinct behaviors depending on the hole position near high-symmetry stacking regions. Peak I is a trapped Wannier type, peak II is trapped with tiny charge-transfer character, and peak III is delocalized with significant charge-transfer character.}
\label{fig2}
\end{figure*}

\begin{figure*}[ht!]
    \centering
    \includegraphics[scale=0.45]{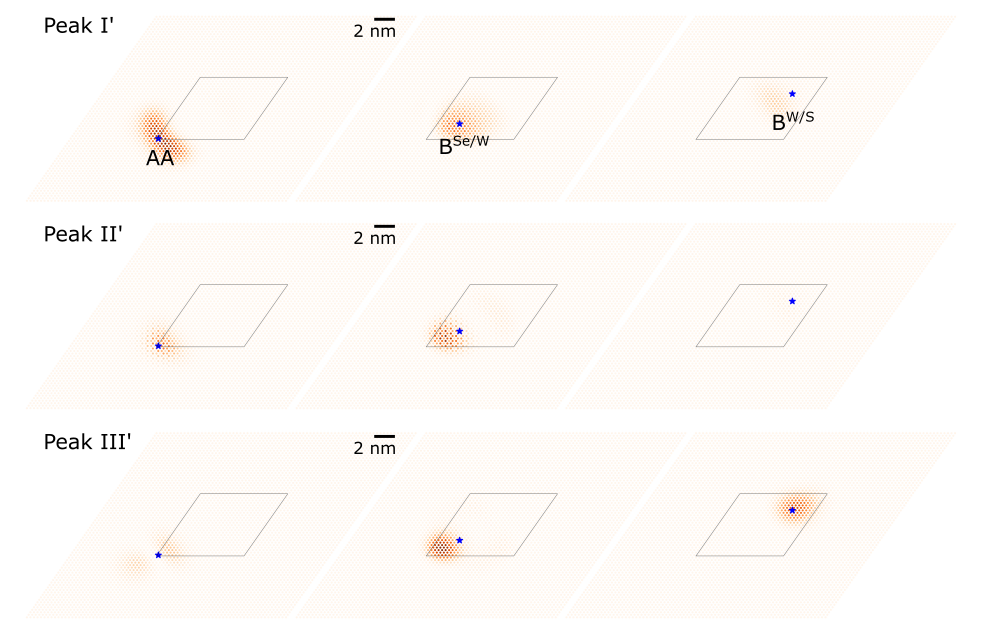}
    \caption{\textbf{Spatial distribution of intralayer exciton wavefunctions
for a fixed hole in parallely aligned WS$_{2}$/WSe$_{2}$ (without hBN
screening)} At $\theta = 0^\circ$, peaks I$^\prime$, II$^\prime$, and
III$^\prime$ (as shown in Fig.~\ref{theoryvsexpt}(c)) exhibit distinct
behaviors depending on the hole position near high-symmetry stacking regions.}
\label{fig2suspended}
\end{figure*}

\textbf{Optical conductivity of intralayer excitons of WSe$_{2}$}:
Figures~\ref{theoryvsexpt}(a),(b) compare the computed optical conductivity of
WSe$_{2}$ in a WS$_{2}$/WSe$_{2}$ heterobilayer with previously observed
reflection contrast spectra with hBN as substrate on both sides~\cite{Jinobservation2019}.
The GW correction is applied as a rigid shift to the monolayer WSe$_2$ excitons (also known as the \textit{scissor operator}~\cite{Leonself2025}). At a large twist
angle, the optical conductivity is dominated by a single peak at about 1.7 eV,
similar to that of monolayer
WSe$_{2}$~\cite{Maityatomistic2025,Aslanexcitons2022,Chensuperior2018,
Stiermagnetooptics2018, Kusubabroadband2021, Liumagneto2019}. See SI, Sec.~A and B, for the monolayer results. This is often referred to
as the ``A" exciton. The optical conductivity of parallelly aligned
WS$_{2}$/WSe$_{2}$ ($\theta=0^\circ$) shows three peaks, labeled I, II, and
III~\cite{Jinobservation2019,Sandhyahyperspectral2022,Kaidistinct2025,Liuexcitonic2021}.
The separations between peaks I-II and II-III are 47 meV and 39 meV, in good
agreement with the experimentally observed values of 56 meV and 37 meV,
respectively. Our computed energies agree better with
experiments~\cite{Jinobservation2019,Sandhyahyperspectral2022,Kaidistinct2025,Liuexcitonic2021}
than previous GW-BSE results~\cite{Naikintralayer2022} and improved continuum
model~\cite{Wangtwist2025}, which do not include screening from hBN
encapsulation. We also observe a redshift of peak I by 11 meV relative to the
dominant A exciton at large twist angles, in qualitative agreement with the
experimentally measured redshift of 22 meV (see
Fig.~\ref{theoryvsexpt}(a),(b)). However, our calculations underestimate the
oscillator strength of peak III. As we focus on intralayer excitons in
WSe$_{2}$ within the heterobilayer, we include the presence of WS$_{2}$ only
through atomic rearrangements and neglect its direct impact on WSe$_{2}$’s
electronic structure (see SI, Sec.~A for details). This may be a source of
underestimation of the oscillator strength of peak III. We note that other
experiments are consistent with our computed oscillator
strengths~\cite{Kaidistinct2025}. 

\begin{figure*}[ht!]
    \centering
    \includegraphics[scale=0.375]{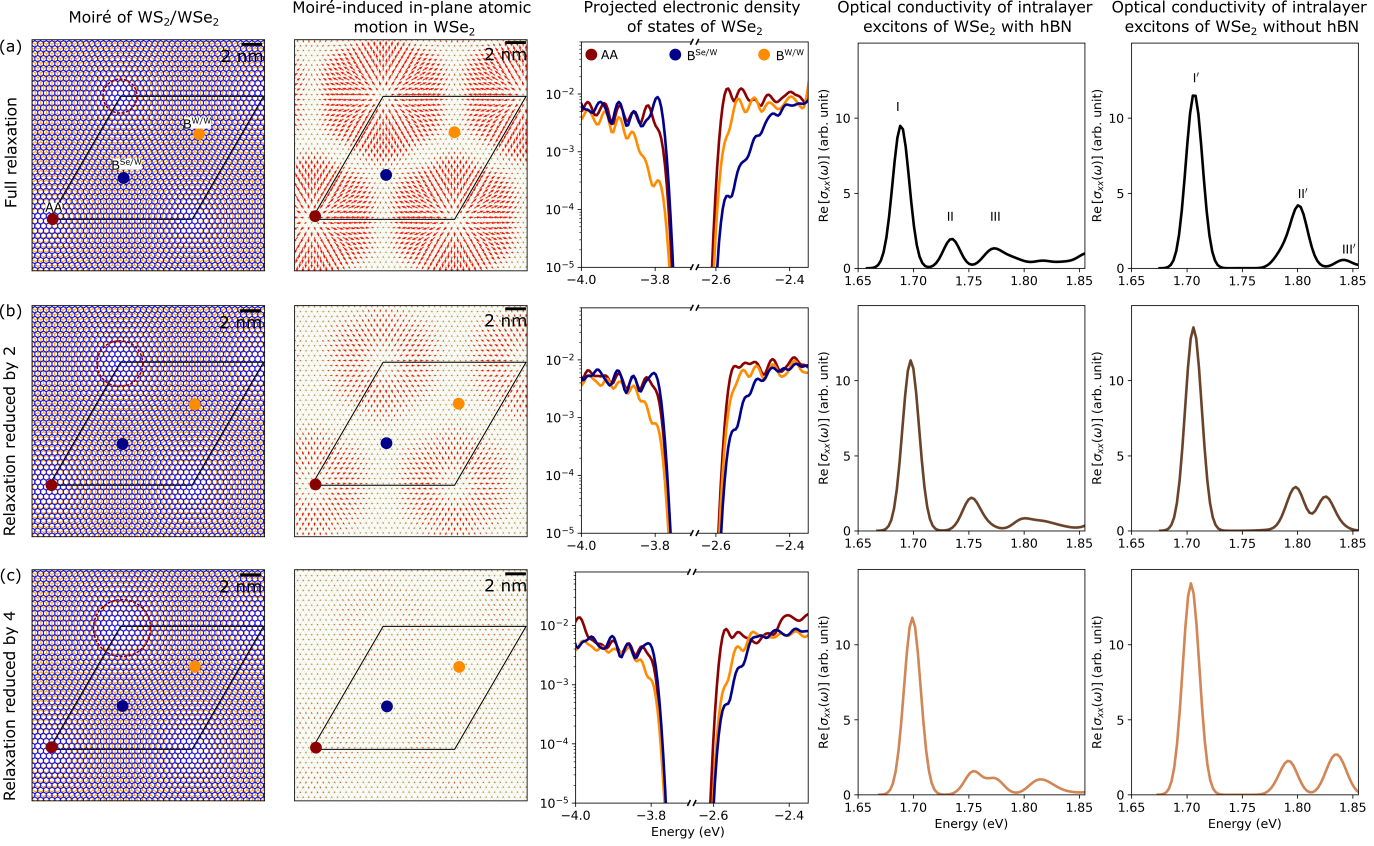}
    \caption{\textbf{\textit{Adiabatic switching} of atomic relaxations and its
impact} Moiré pattern and in-plane atomic displacements, with atomic relaxation
decreasing from top to bottom in the first two columns. The expansion of the AA
region with reduced relaxation is highlighted with a circle. The impact of this
adiabatic switching on electronic and excitonic structures is shown in the next
three columns.} \label{fig4}
\end{figure*}

Fig.~\ref{theoryvsexpt}(c) shows the optical conductivity of the same system in
the absence of environmental screening from hBN. In the aligned
WS$_{2}$/WSe$_{2}$, we find two bright moiré excitons, labeled I$^\prime$ and
II$^\prime$. The third peak, labeled III$^\prime$, exhibits significantly
weaker oscillator strength. Peaks I$^\prime$ and III$^\prime$ are separated by
137 meV, well above the typical experimental value of about 90
meV~\cite{Jinobservation2019,Sandhyahyperspectral2022,Kaidistinct2025,Liuexcitonic2021}. While experiments usually show a redshift of the lowest bright
exciton, our calculation shows a slight blueshift (see
Fig.~\ref{theoryvsexpt}(c)). Our calculations clearly demonstrate that substrate screening is essential to \textit{quantitatively} capture all experimentally observed features of intralayer excitons in WS$_2$/WSe$_2$.

\textbf{Intralayer exciton wavefunctions of WSe$_{2}$}: To examine the spatial
character of bright excitons (as shown in Fig.~\ref{theoryvsexpt}(a)), we plot
the electron density with a fixed hole in the moiré cell near the high-symmetry
stacking regions, as obtained from Eqn.~\ref{exwfn_electron}. 
\begin{equation}
\Big|\Psi^{S}({\bf r}_{e}={\bf R}_{j_{1}} + {\bf t}_{n_{1}}, {\bf r}_{h}^{\mathrm{W}})\Big|^{2} \propto \Big| \sum_{cv{\bf k}} A^{S}_{cv{\bf k}} \times \sum_{n_{3}\in{\bf r}_{h}^{\mathrm{W}}} (C_{n_{3}v}^{\bf k})^{*} \times e^{i{\bf k}.{\bf R}_{j_{1}}} C_{n_{1}c}^{\bf k}\Big|^{2}
\label{exwfn_electron}
\end{equation}
Here, $n_{3}\in{\bf r}_{h}^{\mathrm{W}}$ denotes the Wannier functions centered
at a tungsten atom ${\bf r}_{h}^{\mathrm{W}}$, $A^{S}_{cv{\bf k}}$ is
the BSE eigenvector corresponding to eigenvalue $S$, $C^{\bf k}_{nm}$ are the
coefficients expanding the $m$-th Bloch wavefunction in the Wannier basis (with
$m$ representing either a conduction band $c$ or a valence band $v$), ${\bf R}$
are supercell lattice vectors commensurate with the ${\bf k}$-grid, and
$\mathbf{t}_n$ represents the position of the $n$-th Wannier function in the
home moir\'{e} unit cell. At large twist angle, the A peak resembles a $1s$
hydrogen-like state, with the hole at the center regardless of its position in
the moiré cell (see Fig.~\ref{fig2}(a)). This is a Wannier exciton that is not
trapped in any specific high-symmetry region. The A exciton has a Bohr radius
$a_{\mathrm{B}} = 1.17$ nm, obtained from $\big |\Psi^{\mathrm{S}} ({\bf r} =
{\bf r}_{e}-{\bf r}_{h}^{\mathrm{W}})\big|^{2} \propto
e^{-\frac{r}{a_{\mathrm{B}}}}$. On the other hand, the excitons show a
remarkably different behavior in the parallelly aligned WS$_{2}$/WSe$_{2}$
heterobilayer. For peak I, Fig.~\ref{fig2}(b) shows significant electron
density only when the hole is located in the AA region. Although the electron
density generally ``follows" the hole, contributions from other regions remain
small. Peak I is a trapped Wannier exciton with a Bohr radius of 1.17 nm
localized in the AA region. Our exciton localizations agree well with
stacking-resolved electron energy loss spectroscopy intensity maps around peak
I~\cite{Sandhyahyperspectral2022}. The calculated localization for Peak I agrees with GW-BSE results~\cite{Naikintralayer2022} but differs from improved continuum models~\cite{Wangtwist2025}, both obtained without hBN screening. For peak II, Fig.~\ref{fig2}(b) shows maximum electron density
halfway between the AA and $\mathrm{B^{Se/W}}$ regions when the hole is in the
$\mathrm{B^{Se/W}}$ region. However, the electron density shows non-negligible
overlap when the hole is in the AA region. Peak II is a trapped exciton with a
small charge-transfer character. Fig.~\ref{fig2}(b) shows that for peak III,
the maximum electron density is in the AA region when the hole is in the
$\mathrm{B^{Se/W}}$ region. This indicates strong charge-transfer character. We
also note that when the hole is located in the AA region, the electron density
exhibits hexagon-like distributions with corners at AA sites. Peak III is more
delocalized than the other two peaks, with significant charge-transfer
character (See SI, Sec.~G for more details). While previous GW-BSE results obtained without hBN screening also
indicate charge-transfer character, Peak III is found to be significantly more
localized. For instance, with the hole in the AA region, GW-BSE predicts no
electron density contribution to the Peak III exciton, in contrast to the
hexagon-like distribution found in our calculations. 

Fig.~\ref{fig2suspended} shows the electron densities with a fixed hole in the moiré cell near the high-symmetry stacking regions, without hBN screening. Peak I$^\prime$ shows the largest electron density when the hole is at the AA region. However, contributions from the $\mathrm{B^{Se/W}}$ regions are non-negligible. Peak I$^\prime$ is not as strongly trapped as Peak I. Overall, peaks I$^\prime$ and II$^\prime$ exhibit features qualitatively similar to those of peaks I and II. In contrast, Peak III$^\prime$ exhibits qualitatively distinct behavior from the Peak III exciton wavefunction, showing no charge-transfer character and significant contributions from $\mathrm{B^{W/S}}$ regions. In this context, peak III from previous GW-BSE is best compared to peak $IV^\prime$ in our results (see SI, Sec.~XX). Our calculations clearly show that the inclusion of substrate screening \textit{qualitatively} modifies the characteristics of intralayer excitons in WS$_2$/WSe$_2$.

\textbf{Origin of moir\'{e} trapping and charge-transfer in intralayer
excitons}: To gain atomistic insight into moir\'{e} trapping of excitons, we
employ an \textit{adiabatic switching} approach: the atomic relaxation patterns
of the fully relaxed WS$_{2}$/WSe$_{2}$ are systematically reduced in
amplitude, and intralayer excitons are computed at each step. Among the
high-symmetry stacking regions, AA is the most energetically unfavorable. In
the AA region, atoms move radially outward to reduce energy, as shown by the
in-plane displacements in Fig.~\ref{fig4}. The displacement is largest in the
fully relaxed structure. The AA regions form \textit{aster} topological defects
with charge +1~\cite{Anghelutatopological2025}. As a result, the electron
density of states projected onto the high-symmetry stacking regions shows
significant changes near the electronic band gap. These changes are calculated
using density functional theory and are shown in
Fig.~\ref{fig4}. In the fully relaxed calculation, the valence band edge mainly
comes from $\mathrm{B^{Se/W}}$, with AA regions close behind. The conduction
band edge, in contrast, mainly comes from AA regions, with $\mathrm{B^{W/S}}$
close behind. Reducing the relaxation by a quarter makes the valence band
contributions from $\mathrm{B^{Se/W}}$ and AA nearly equal, with
$\mathrm{B^{W/S}}$ close behind. For the conduction band, the balance between
$\mathrm{B^{W/S}}$ and $\mathrm{B^{Se/W}}$ shifts, but the AA regions still
dominate the edge. 

The intricate interplay of local density-of-states orderings across these
regions is crucial for moiré trapping and the origin of excitons with charge
transfer charcater. We first demonstrate this by performing calculations with
hBN screening. As relaxation decreases, the lowest bright excitons show two
main changes (fourth column of Fig.~\ref{fig4}). First, the spectrum blueshifts
by 11 meV, with the lowest-energy bright exciton aligning to the A exciton at
large twist angles (see Fig.~\ref{fig4} and Fig.~\ref{theoryvsexpt}). Second,
the energy separation between I and III increases from 86 meV to 116 meV.
Specifically, the oscillator strength of peak I increases by 24$\%$ as
relaxation is reduced to one quarter. This increase is a direct consequence of
moir\'{e} detrapping. Qualitatively, the oscillator strength of peak I is
proportional to $\sum_{n_{1},n_{2}} \big(C^{\bf k}_{n_{1}c}\big)^{*} C^{\bf
k}_{n_{2}v}$. In the detrapped case, Wannier
functions at ${\bf t}_{n}$ from all high-symmetry stacking regions contribute.
By contrast, in the trapped case, only the AA region contributes. Exciton
wavefunctions confirming this behavior are shown in SI, Sec.~D. Moreover, the
oscillator strength of peak II relative to peak I,
$\frac{\sigma_{xx}^{II}}{\sigma_{xx}^{I}}$, decreases by 53$\%$ as relaxation
is reduced to one quarter. None of the bright excitons of experimental interest
(up to 100 meV above the lowest bright exciton) exhibit charge-transfer
character when the relaxation is reduced to one quarter (see SI, Sec.~D for
details). The intricate interplay between atomic structure and exciton spectra
found in our adiabatic switching simulations may explain the variable exciton
confinement observed in previous scanning transmission electron microscopy -
electron energy loss spectroscopic
measurements~\cite{Sandhyahyperspectral2022,Dandumapping2024,Sankarseeing2025}.
These measurements spectrally average over 100 moir\'{e} unit cells, smoothing
out the nanoscale structural inhomogeneities common in moiré
materials~\cite{Deadvanced2025}. 

The last column of Fig.~\ref{fig4} shows the system’s optical conductivity without hBN screening in the electron–hole interactions. The lowest
bright exciton energy is relatively insensitive to the degree of atomic
relaxation and shows a redshift as relaxation decreases. The separation between
the three peaks labeled I$^\prime$, II$^\prime$, and III$^\prime$ remains
nearly constant at about 135 meV, regardless of the degree of relaxation. The
oscillator strength of peak III$^\prime$ increases significantly as relaxation
decreases. In particular, the relative peak intensities appear deceptively
similar to those observed in experiments (see the last column of
Fig.~\ref{fig4}(c) and Fig.~\ref{theoryvsexpt}(b)). We analyze the exciton wave
functions associated with the peaks and find that none exhibit significant
charge-transfer character (see SI, Sec.~E). Peaks I$^\prime$ and
II$^\prime$ remain largely detrapped, while peak III$^\prime$ localizes on the
AA region. Our adiabatic switching simulations establishes that atomic-relaxation-driven direct–indirect band-gap transitions, together with electron–hole interactions tunable by environmental screening, give rise to rich features of moir\'{e} intralayer excitons.

\begin{figure*}[ht!]
    \centering
    \includegraphics[scale=0.37]{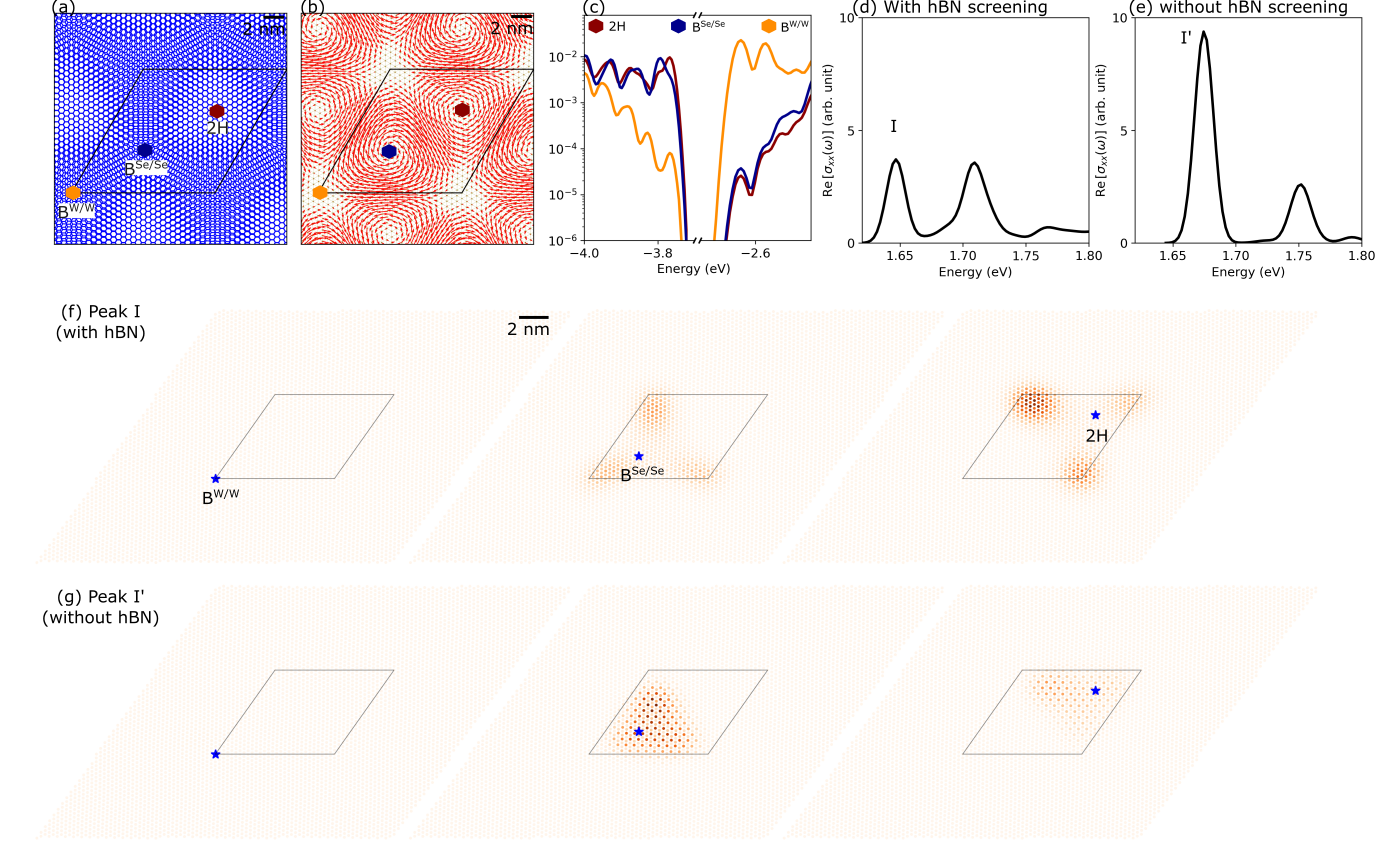}
    \caption{\textbf{Charge-transfer character of the lowest-energy bright
exciton in 57.7$^\circ$ twisted bilayer WSe$_{2}$}. (a) Atomic structure of the
reconstructed moir\'{e} cell with the high-symmetry stacking regions marked.
Corresponding impact on bottom WSe$_{2}$ layer: (b) in-plane atomic
displacements induced by the moir\'{e}, (c) stacking-resolved local electron
density of states. (d), (f) and (e), (g): Optical conductivity and exciton
wavefunction for lowest bright exciton, with the hole fixed at a high-symmetry
stacking region (blue star), shown with and without hBN screening,
respectively.} \label{fig5}
\end{figure*}

\subsection{Intralayer excitons of WSe$_{2}$ for WSe$_{2}$/WSe$_{2}$ homobilayer}
Regardless of the competition with charge-transfer character, the lowest-energy
bright intralayer exciton (peak I) in WSe$_{2}$ of WS$_{2}$/WSe$_{2}$ is always
Wannier type. We show that this ordering can be manipulated. Fig.~\ref{fig5}
shows an example in twisted bilayer WSe$_2$ with a moir\'{e} size nearly
identical to that of aligned WS$_2$/WSe$_2$. Fig.~\ref{fig5}(a) shows that
atomic relaxations generate three-fold symmetry around the energetically
unfavorable $\mathrm{B^{Se/Se}}$ region. The most favorable 2H region (W above
Se and Se above W) occupies the largest area of the moiré cell, consistent with
experimental observations~\cite{Lilattice2021,Westonatomic2020} and
theory~\cite{Maityreconstruction2021,Enaldievstacking2020,Carrrelaxation2018}.
In-plane atomic displacements are tangential around $\mathrm{B^{Se/Se}}$,
unlike the radial pattern in WS$_{2}$/WSe$_{2}$ (see Fig.~\ref{fig5}(b)). The
$\mathrm{B^{Se/Se}}$ regions form \textit{vortex} topological defect with
charge +1~\cite{Anghelutatopological2025}.  Fig.~\ref{fig5}(c) shows the
stacking-resolved local electronic density of states near the band gap. The
band gap remains spatially indirect by a substantial margin.
Figs.~\ref{fig5}(d),(f) demonstrate the charge-transfer character of peak I,
the lowest-energy bright exciton. Peak I shows a 56 meV redshift and about a 72
\% decrease in oscillator strength compared to the A peak of monolayer
WSe$_{2}$. Figs.~\ref{fig5}(e) and (g) show the optical conductivity and
exciton wavefunctions without hBN screening. Remarkably, we show that Peak
I$^\prime$ is of Wannier type when hBN encapsulation screening is excluded. 

\subsection{Discussions} 
The interplay between intralayer moiré charge-transfer and Wannier-like
excitons emerges naturally when considering direct interactions in the
Wannier-function basis (see SI, Sec.~A). The direct
interaction can be viewed as a product of two components, $|C_{n_{1}c}^{\bf
k}|^{2}|C_{n_{3}v}^{\bf k}|^{2}$ and $W({\bf t}_{n_{3}}-{\bf t}_{n_{1}})$. The
larger this product, the higher the exciton binding energy. At short distances
relevant to Wannier exciton, $W \sim
-\frac{1}{r_{0}}\ln\big(\frac{r\epsilon_{\mathrm{env}}}{r_{0}}\big)$. At large
distances relevant to charge-transfer exciton, $W\sim
\frac{1}{r\epsilon_{\mathrm{env}}}$. We find that $W$ binds the Wannier exciton
about nine times more strongly than the charge-transfer exciton, using $r_0 =
4.5~\mathrm{nm}$, $\epsilon_{\mathrm{env}} = 3.8$, $r = 0.1~\mathrm{nm}$ for
the Wannier exciton, and $5~\mathrm{nm}$ for the charge-transfer exciton. The
charge-transfer exciton is favored only when $|C_{n_{1}c}^{\bf k}|^{2}
|C_{n_{3}v}^{\bf k}|^{2}$ is very small for ${\bf t}_{n_{3}} - {\bf t}_{n_{1}}
\approx 0$. Our calculations demonstrate that the character of \textit{intralayer excitons}, from Wannier-like to charge-transfer-like, can be tuned by environmental screening (through $\epsilon_{\mathrm{env}}$) and by moiré relaxation engineered via stacking or twisting. Wannier-like and
charge-transfer-like trapped states can be probed directly using photocurrent
tunneling microscopy~\cite{Liimaging2024} or indirectly through
doping-dependent absorption spectra~\cite{Naikintralayer2022}. Although Coulomb engineering in layered materials has been widely studied~\cite{Rajacoulomb2017}, its interplay with moiré relaxation in shaping exciton characteristics remains underexplored. In this context, our results should motivate the exploration of other excitonic species, including layer-separated dipolar and quadrupolar excitons.

In summary, we establish an accurate and computationally feasible atomistic many-body approach to study excitons in moiré materials. Our results pave the way for atomistic design of excitonic states in van der Waals heterostructures and the exploration of rich many-body physics mediated by exciton–phonon–light interactions.  

\begin{acknowledgement}
This work used the ARCHER2 UK National Supercomputing Service via our
membership of the UK Car-Parrinello Consortium, which is funded by the EPSRC grant EP/X035891/1. We thank Emma Regan, Chenhao Jin, and Feng Wang for providing the experimental data.
\end{acknowledgement}


\clearpage
\newpage 

\section{A. Simulation Methods and Results for Monolayer WSe$_{2}$}
\textbf{Atomic structure} All twisted WS$_{2}$/WSe$_{2}$ and WSe$_{2}$/WSe$_{2}$ moir\'{e} structures were generated using the TWISTER package~\cite{Naiktwister2022}. We used lattice constants of 3.32~$\mathrm{\AA}$ for monolayer WSe$_{2}$ and 3.18~$\mathrm{\AA}$ for WS$_{2}$. Atomic relaxations were captured using classical interatomic potentials fitted to data obtained from density functional theory. We used the Stillinger-Weber potential for interactions within each layer~\cite{Zhouhandbook2017} and the Kolmogorov-Crespi potential for interactions between layers~\cite{Naikkolmogorov2019}. 
All relaxations were performed using the LAMMPS package (https://lammps.org/)~\cite{lammps}. We used the FIRE algorithm~\cite{Bitzekstructural2006} to relax the atoms within a fixed simulation box, applying a force tolerance of $10^{-4}$eV/$\mathrm{\AA}$ for each atom in all directions. We reduced the atomic relaxations to half and one-quarter of the full relaxation relative to the unrelaxed structures to obtain the configurations used for \textit{adiabatic switching}. The 16.1$^\circ$ and parallelly aligned WS$_{2}$/WSe$_{2}$ heterobilayers contain 75 and 3903 atoms in the moir\'{e} unit cell, respectively. The 57.72$^\circ$ twisted WSe$_{2}$/WSe$_{2}$ homobilayer contains 3786 atoms in its moir\'{e} unit cell.

\textbf{Electronic structure} Electronic structure calculations were performed using the SIESTA package~\cite{Solersiesta2002}, which employs localized atomic orbitals as the basis. We employed norm-conserving Troullier-Martins pseudopotentials~\cite{Troullierefficient1991} and the local density approximation to describe exchange-correlation effects~\cite{Perdewself1981}.
We used a single-$\zeta$ plus polarization basis before expanding the wavefunctions in the Wannier basis, which has been shown to be sufficient for capturing low-energy moiré effects~\cite{Maityatomistic2025}. Using a double-$\zeta$ plus polarization basis as well as a plane-wave basis to perform electronic structure calculations before generating Wannier functions results in almost identical bright excitons in monolayer WSe$_2$~\cite{Maityatomistic2025}. For small twist angles we used the $\Gamma$ point and for large twist angles we employed a $3\times3\times1$ $k$-grid to sample the moir\'{e} Brillouin zone for electronic charge density calculations. A plane-wave energy cutoff of 100 Rydberg was used. We used an energy shift of 0.02 Ry. A vacuum spacing of 20 $\mathrm{\AA}$ was applied in the out-of-plane direction. 

The electronic density of states was calculated with a fine $3 \times 3 \times 1$ $k$-grid of the moir\'{e} Brillouin zone, corresponding approximately to a $78 \times 78 \times 1$ $k$-grid of the monolayer WSe$_{2}$ Brillouin zone. We used a 30~meV Gaussian to represent the $\delta$ function in the density of states calculation. To obtain stacking-resolved local density of states, we average atomic orbital contributions to the density of states over a circular in-plane patch of 9~$\mathrm{\AA}$ radius around the center of each selected high-symmetry stacking region~\cite{Molinoinfluence2023}. Spin-orbit coupling was included only in the projected density of states calculations~\cite{Fernandezonsite2007}.

\textbf{Wannier functions} We followed the strategy of generating Wannier functions to expand the electronic wavefunctions, as detailed in our previous work~\cite{Maityatomistic2025}. We used the one-shot projection method~\cite{Marzarimaximally2012} in the Wannier90 code~\cite{Pizziwannier902020} to construct the Wannier functions. The input files for Wannier90, including overlaps between projection orbitals and Bloch states at each $k$-point and band (the so-called \texttt{amn} file) and overlap matrices between cell-periodic parts of Bloch states at neighboring k-points (the so-called \texttt{mmn} file), were generated using SIESTA. A disentanglement procedure was used~\cite{Marzarimaximally2012,Souzamaximally2001}. Our previous work~\cite{Maityatomistic2025} showed that electronic structure calculations using Wannier functions as a basis agree excellently with SIESTA results obtained without Wannierization. In this work, we used only the $\Gamma$ point of the moir\'{e} Brillouin zone to construct the Hamiltonian with the Wannier basis in real space. We then diagonalize this Hamiltonian on the required $k$-grid to obtain the electronic structure. This approach is conceptually the same as Wannier interpolation~\cite{Pizziwannier902020} and significantly reduces memory usage and enables efficient handling of larger systems than previously possible.

Wannier Hamiltonian in real space can be written as
\begin{equation}
\tilde{H}_{mn{\bf R}} = \langle w_{m{\bf 0}}| H | w_{n{\bf R}}\rangle 
\end{equation}
where, $w_{n{\bf R}}$ is the Wannier function for the $n$-th band, and ${\bf R}$ denotes the lattice vectors. We employed an improved interpolation scheme, based on minimal-distance replica selection, to Fourier transform the Wannier Hamiltonian~\cite{Pizziwannier902020}. This is outlined below
\begin{equation}
H_{mn{\bf k}} = \sum_{\bf R} \frac{1}{\mathcal{N}_{mn{\bf R}}} \sum_{j=1}^{\mathcal{N}_{mn{\bf R}}} e^{i{\bf k}.({\bf R}+ {\bf T}^{j}_{mn{\bf R}})}\tilde{H}_{mn{\bf R}}
\label{FTImproved}
\end{equation}
where ${\bf T}^{j}_{mn{\bf R}}$ are $\mathcal{N}_{mn{\bf R}}$ vectors ${\bf T}$ that minimizes distance $|{\bf r}_{m} - ({\bf r}_{n} + {\bf R} + {\bf T})|$ for a given combination of $m$, $n$, and ${\bf R}$. They are directly available from the Wannier90 output files, namely \texttt{*hr.dat} and \texttt{*wsvec.dat}. We obtained eigenvalues and eigenvectors by diagonalizing the Fourier-transformed Hamiltonian, and also computed $\nabla H_{k}$, which is required for calculating dipole operators. To diagonalize, we use NumPy’s LAPACK routines for small systems and the ELPA libraries for selected large matrices~\cite{Auckenthalerparallel2011, MarekElpa2014}.

\textbf{Exciton calculations} All exciton calculations were performed using an improved version of PyMEX package, a Python package for Moir\'{e} EXcitons that solves the Bethe–Salpeter Equation. The previous version of the code is publicly available on GitHub at \url{https://github.com/imaitygit/PyMEX}. The new version of the code, with additional features, better scalability, and documentation will be released in the future. We summarize the key equations relevant to this work, with full derivations available elsewhere~\cite{Maityatomistic2025}. The BSE Hamiltonian for zero-momentum excitons can be approximated as, 
\begin{equation}
\begin{split}
& \langle cv{\bf k}| \hat{H}_{\mathrm{BSE}} |c^{\prime}v^{\prime}{\bf k^{\prime}}\rangle \\
& \approx (\epsilon_{c{\bf k}} - \epsilon_{v{\bf k}}) \delta_{cc^\prime} \delta_{vv^\prime} \delta_{\bf kk^\prime} \\
& -\frac{1}{N} \sum_{{\bf R},n_{1},n_{3}} {\big(C^{\bf k}_{n_{1}c}\big)}^* C^{\bf k^\prime}_{n_{1}c^\prime} {C^{\bf k}_{n_{3}v}} {\big(C^{\bf k^\prime}_{n_{3}v^\prime}\big)}^* W({\bf R} + ({\bf t}_{n_3} - {\bf t}_{n_1})) e^{i ({\bf k - k^\prime}).{\bf R}} \\
& + \frac{1}{N} \sum_{{\bf R},n_{1},n_{3}} {\big(C^{\bf k}_{n_{1}c}\big)}^* {C^{\bf k}_{n_{1}v}} C^{\bf k^\prime}_{n_{3}c^\prime} {\big(C^{\bf k^\prime}_{n_{3}v^\prime}\big)}^* V({\bf R} + ({\bf t}_{n_{3}} - {\bf t}_{n_{1}}))
\label{bsewannier}
\end{split}
\end{equation}
where $\epsilon_{c(v)\mathbf{k}}$ denotes the quasiparticle energy of an electron in the conduction (valence) band $c$ ($v$) with crystal momentum $\mathbf{k}$, $\mathbf{R}$ denotes lattice vectors, $C_{nm}^{\mathbf{k}}$ are the expansion coefficients for electron wavefunctions in the Wannier basis, $\mathbf{t}_n$ represents the position of the $n$-th Wannier function in the home unit cell, $N$ is the number of $k$ points used to sample the moir\'{e} Brillouin zone, and $W$ and $V$ are the screened Coulomb and bare Coulomb potentials, respectively. We used the Rytova–Keldysh potential to model the screened Coulomb interaction~\cite{Cudazzodielectric2011,Keldyshcoulomb1979,Berkelbachtheory2013,Chernikovexciton2014,Rytovascreened2018,Vancoulomb2018,Gorycarevealing2019},
\begin{equation}
W({\bf r}) = - \frac{1}{4\pi\epsilon_{0}} \frac{\pi e^{2}}{2r_{0}} \Bigg[ H_{0}\Big( \frac{\epsilon_{\mathrm{env}}r}{r_{0}}\Big) -  Y_{0}\Big( \frac{\epsilon_{\mathrm{env}}r}{r_{0}}\Big) \Bigg]
\end{equation}
where $r_{0}$ is the screening length in the absence of a substrate, $\epsilon_{\mathrm{env}}$ is the effective dielectric constant because of the substrate, and H$_{0}$ and Y$_{0}$ are the zeroth-order Struve and Bessel functions of the second kind, respectively. Importantly, all of these parameters were directly obtained from ab initio calculations. For monolayer WSe$_{2}$, $r_0$ is set to 45~$\mathrm{\AA}$~\cite{Berkelbachtheory2013}, and $\epsilon_{\mathrm{env}} = \frac{(\epsilon_t + \epsilon_b)}{2} = 3.8$~\cite{Laturiadielectric2018}, corresponding to hexagonal boron nitride encapsulation on both sides. With these parameters, we captured the $1s$ exciton binding energy and the separation in energy between the $2s$ and $1s$ states within a few meV of most experimental results on monolayer WSe$_{2}$ (see SI, Sec.~B for details). The GW correction can be applied as a rigid shift to the monolayer WSe$_2$ exciton peak~\cite{Maityatomistic2025}. Although such corrections could be applied with hBN encapsulation, we did not perform them. Instead, we aligned the monolayer WSe$_2$ A peak ($1s$ exciton) to 1.7 eV. We emphasize that no additional free parameters were introduced in our simulations. Benedict~\cite{Screeningbenedict2002} argued that, when constructing the BSE Hamiltonian using a finite set of electron-hole states, it is necessary to screen the bare exchange Coulomb integral, which is generally unscreened~\cite{Besterelectronic2009,Qiusolving2021}. This is known as the S-approximation. All results in the main text use the S-approximation, as we construct the BSE Hamiltonian with only a small subset of valence and conduction bands. In the SI, Sec.~C, we compare calculations performed with screened and unscreened exchange integrals and find no significant differences.

The optical conductivity of excitons was computed using~\cite{Maityatomistic2025,Pedersonoptical2001},
\begin{equation}
\text{Re}[\sigma_{xx}(\omega)] \propto \frac{1}{\hbar \omega}\sum_S \left| \sum_{cv\mathbf{k}} A_{cv\mathbf{k}}^S \langle v\mathbf{k} | p_x | c\mathbf{k} \rangle \right|^2 \delta(\omega - \omega^S)
\end{equation}
Here, $A_{cv\mathbf{k}}^S$ are the exciton eigenvectors corresponding to the excitonic state $S$, and $\langle v\mathbf{k} \,|\, p_x \,|\, c\mathbf{k} \rangle$ represents the matrix element of the momentum operator between the valence ($v$) and conduction ($c$) band states. In all our calculations, the delta function was replaced by a Gaussian with a standard deviation of 7.5 meV. The momentum (or, dipole) operator was computed using 
\begin{equation}
\langle v{\bf k}|{\bf p}| c{\bf k}\rangle = \sum_{n_{1},n_{2}} \big(C^{\bf k}_{n_{1}c}\big)^{*} C^{\bf k}_{n_{2}v} \big [\nabla_{\bf k}H^{\bf k}_{n_{1}n_{2}} + i({\bf t}_{n_{2}}-{\bf t}_{n_1})H^{\bf k}_{n_{1}n_{2}}\big]
\end{equation}
Using our new interpolation approach, we compute the momentum operator. using Eq.~\ref{FTImproved}. As the material has in-plane isotropy, we found $\text{Re}[\sigma_{xx}(\omega)] \simeq \text{Re}[\sigma_{yy}(\omega)]$ (see SI, Sec.~F for details). 

In our calculations of intralayer excitons of WSe$_{2}$, we considered (i) parallely aligned WS$_2$/WSe$_2$ with a moir\'{e} cell of 82.9~$\mathrm{\AA}$, 18 moir\'{e} valence and conduction bands each, (ii) 57.7$^\circ$ twisted bilayer WSe$_2$ with a moir\'{e} cell of 83.4~$\mathrm{\AA}$ using a $3\times3\times1$ k-grid, 18 moir\'{e} valence and conduction bands each, and (iii) 16.1$^\circ$ twisted WS$_2$/WSe$_2$ with a moir\'{e} cell of 11.5 ~$\mathrm{\AA}$ using a $21\times21\times1$ k-grid, 4 moir\'{e} valence and conduction bands each. The low-energy valence and conduction bands of WSe$_2$ in both WS$_2$/WSe$_2$ and anti-parallel twisted bilayer WSe$_2$ originate from the $K$ and $K^\prime$ valleys of the monolayer WSe$_2$. The electron wavefunctions of these valleys mainly originate from the $d$-orbitals of W atoms, which are only weakly influenced by interlayer hybridization. After relaxing the full WS$_2$/WSe$_2$ and twisted bilayer WSe$_2$ systems, we isolated the fully relaxed WSe$_2$ and examined its electronic states near the band gap. These states agree well with those obtained from calculations on the full WS$_2$/WSe$_2$ and twisted bilayer WSe$_2$ systems, as shown previously~\cite{Naikintralayer2022,Kundumoire2022,Maityatomistic2025}. This approach is motivated primarily by computational simplicity, rather than by any limitations of the method. Furthermore, spin-orbit coupling can be included perturbatively, producing two identical series in WSe$_2$ monolayers A and B (each corresponding to the spin-split partners of the valence bands), separated by roughly 400 meV~\cite{Maityatomistic2025,Qiuscreening2016}. As we focus on low-energy bright excitons within energy range of about 150 meV, we do not include spin-orbit coupling in our BSE. 

To determine the shift in the absolute energy of the lowest bright exciton, we perform BSE calculations using exactly the same number of valence and conduction bands as in the flat monolayer WSe$_{2}$, with identical lattice constants. This approach allows us to accurately capture the energy shifts without requiring full convergence of the lowest excitons, which would demand a significantly larger number of bands.

\section{A. K-point convergence of monolayer WSe$_{2}$ excitons}
\begin{figure}[ht!]
    \centering
    \includegraphics[scale=1]{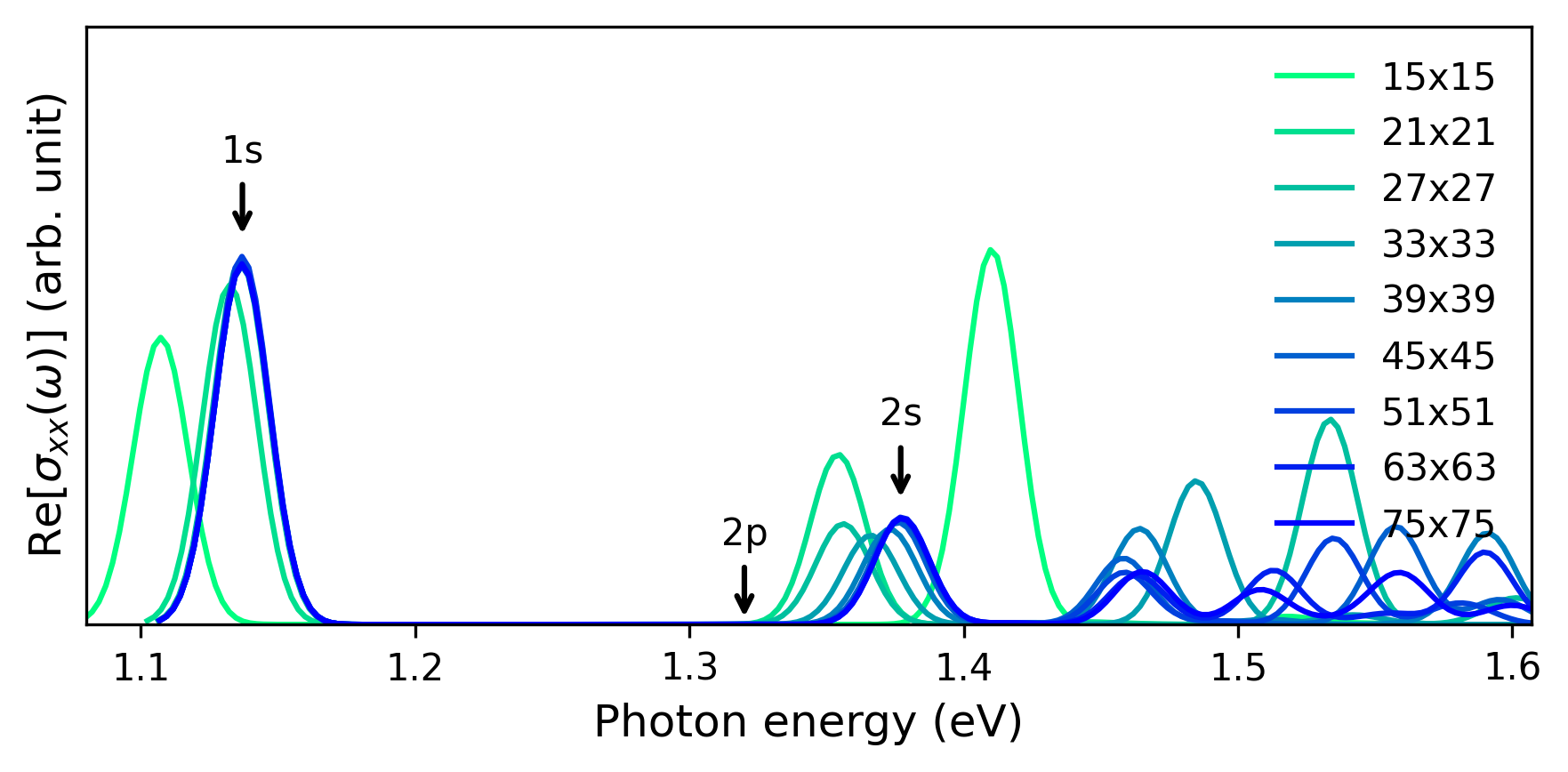}
    \caption{Optical conductivity results for suspended monolayer WSe$_{2}$, computed using electronic wave functions expanded in the Wannier function basis with the PyMEX package. We did not include spin-orbit coupling effects and GW corrections. To highlight the convergence behavior of different excitons, we varied the k-grid used in constructing the Bethe-Salpeter Equation Hamiltonian with one valence and one conduction band. We observe that the $1s$ exciton converges quickly with a $27\times27\times1$ k-grid, the $2p$ exciton with a $45\times45\times1$ k-grid, and the $2s$ exciton with a $51\times51\times1$ k-grid, reflecting the real-space extent of the exciton wavefunctions. The oscillator strength of the $2p$ exciton relative to the $1s$ exciton is about 2,000 times weaker, and that of the $2s$ state is about 4 times weaker. These observations agree well with previous ab initio GW-BSE calculations~\cite{Yeprobing2014}.}
\end{figure}

\clearpage
\newpage 

\section{B. Impact of substrate on monolayer WSe$_{2}$ excitons}
\begin{table}[h!]
\label{Excitonsubstrate}
\caption{Experimentally observed binding energy of the lowest exciton and the energy separation between the two lowest exciton $s$-type Rydberg series, with and without substrates, for monolayer WSe$_2$. In our calculations, we use a static dielectric constant of 3.8 for hexagonal boron nitride (hBN)~\cite{Laturiadielectric2018} and 3.9 for SiO$_{2}$~\cite{Chernikovexciton2014}. In case of hBN, we set $\epsilon_{\mathrm{env}}$ = ($\epsilon_{\mathrm{top}} + \epsilon_{\mathrm{bottom}})/2$ and in case of SiO$_{2}$, we set $\epsilon_{\mathrm{env}} = (\epsilon_{\mathrm{top}} + 1)/2$ to modify the Rytova-Keldysh potential 
with $r_{0}=45$ \AA, consistent with previous ab initio calculations~\cite{Berkelbachtheory2013}. When we include hBN as a substrate, our calculations faithfully reproduce the exciton binding energies observed in most experiments. In monolayer calculations we do not screen the bare Coulomb interaction.}
\begin{tabular}{c|c c c c|c c c c|}
 &  & Binding energy ($1s$) & in eV & & & $\Delta E_{12} \ (E_{2s}-E_{1s})$& in eV  & \\ 
 & Susp. & hBN & SiO$_{2}$ & PDMS & Susp. & hBN & SiO$_{2}$ & PDMS \\
\hline 
Expt.~\cite{Measurementhanbicki2015} & & & 0.89 & & & & 0.79 & \\ 
\hline 
Expt.~\cite{Strainaslan2018} & & & & 0.315 & & & & 0.157 \\ 
\hline 
Expt.~\cite{Aslanexcitons2022} & 0.412 & & & & 0.209 & & &  \\
\hline 
Expt.~\cite{Chensuperior2018}, & & & & & &  0.13 & &  \\
\cite{Stiermagnetooptics2018}, ~\cite{Kusubabroadband2021}, ~\cite{Liumagneto2019}  & & 0.172 &  & & &   & &  \\
\hline 
PyMEX  & 0.471 &  &  & &  0.24  & & &  \\ 
 & & 0.178  &  & &  & 0.125  & & \\ 
 & & (both) &  & &  & (both) & & \\ 
 & & & 0.26 & &  &  & 0.168 & \\
 & & & (top)& &  &  & (top) & \\
\hline 
\end{tabular}
\end{table}

\clearpage
\newpage 

\section{C. Validity of the S approximation~\cite{Screeningbenedict2002}}
\begin{figure}[ht!]
    \centering
    \includegraphics[scale=1]{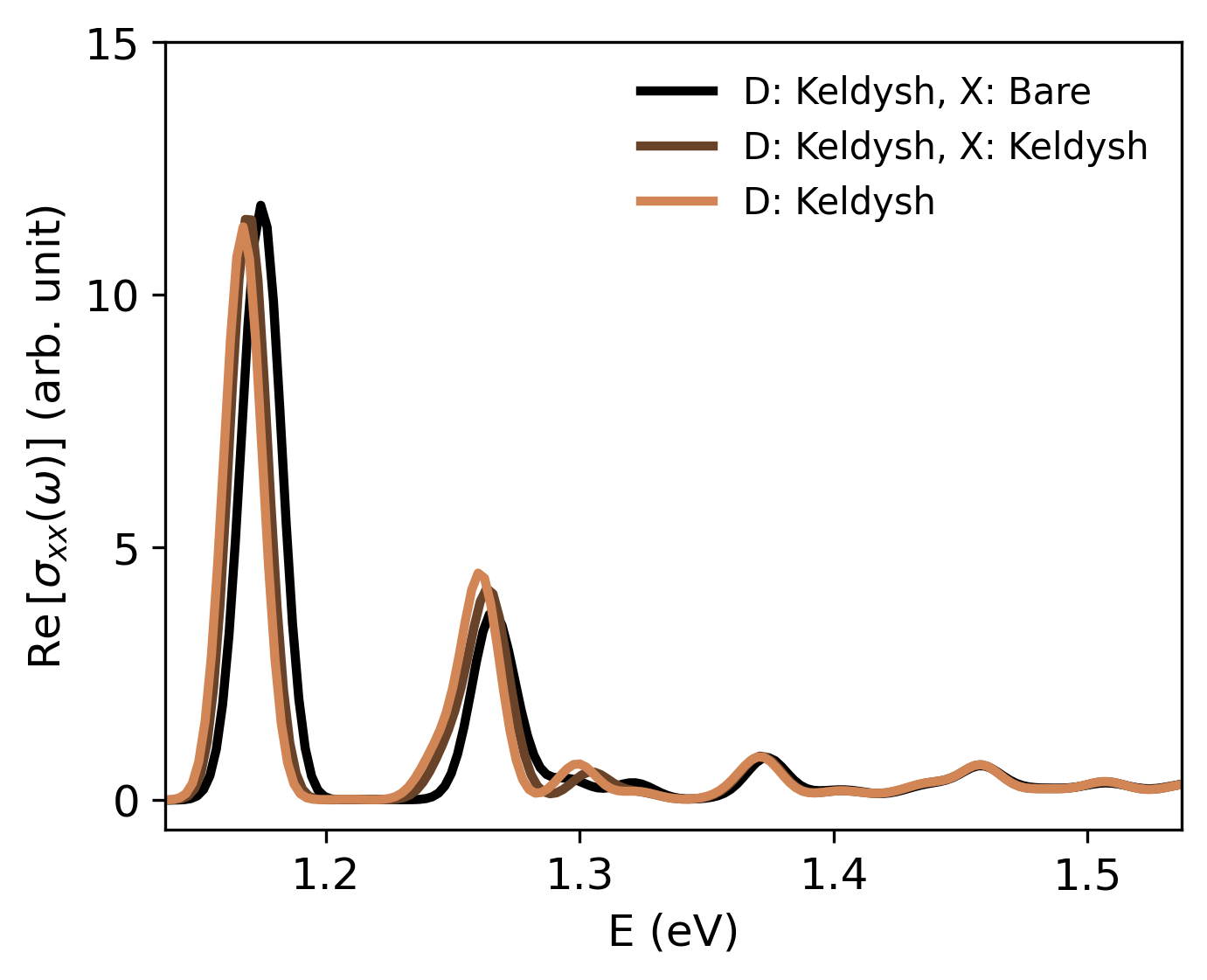}
    \caption{Benedict~\cite{Screeningbenedict2002} argued that, when constructing the BSE Hamiltonian using a finite set of electron-hole states, it is necessary to screen the exchange Coulomb integral, which is generally unscreened. This is the so called S-approximation. Above figure compares the optical conductivity associated with intralayer excitons in suspended WSe$_{2}$ when the BSE Hamiltonian is constructed with the exchange term using the bare Coulomb interaction, the screened Keldysh potential, or by neglecting the exchange term altogether. In all our calculations, we included 18 moiré valence and conduction bands to construct the BSE Hamiltonian. Our results remain consistent across all cases.}
\end{figure}


\clearpage
\newpage

\section{D. Exciton wavefunctions with relaxation reduced to one quarter in parallely aligned WS$_{2}$/WSe$_{2}$}
\begin{figure}[ht!]
    \centering
    \includegraphics[scale=0.5]{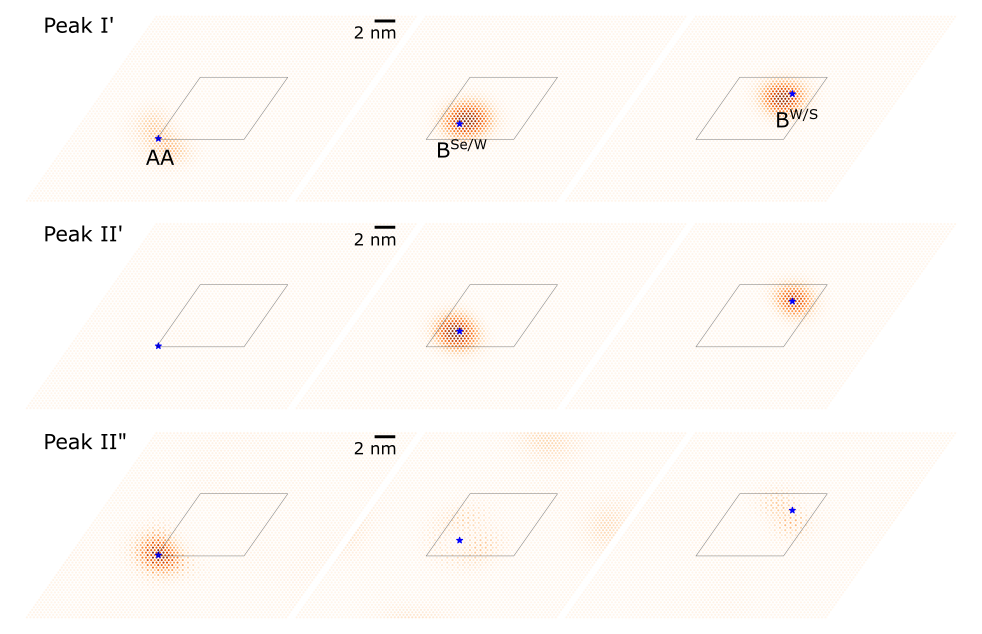}
    \caption{Spatial distribution of intralayer exciton wavefunctions of WSe$_{2}$ for a fixed hole with relaxation reduced to one quarter in parallely aligned WS$_{2}$/WSe$_{2}$ (with hBN screening). We label the three peaks as I$^\prime$, II$^\prime$, and II$^{\prime\prime}$, which are separated by 75 meV. Peaks I$^\prime$ and II$^\prime$ remain largely detrapped, while peak II$^{\prime\prime}$ localizes on the AA stacking region. None of the peaks exhibit charge-transfer character. We show only the exciton wavefunctions of bright excitons within 100 meV.}
\end{figure}

\clearpage
\newpage

\section{E. Impact of adiabatic switching on the intralayer moir\'{e} excitons without environmental screening}
\begin{figure}[ht!]
    \centering
    \includegraphics[scale=0.5]{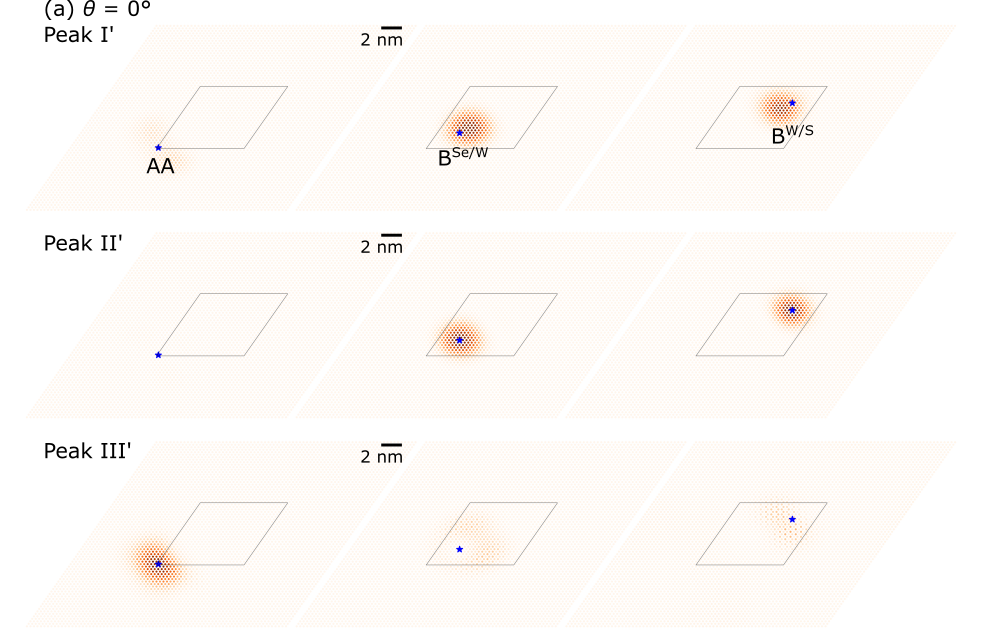}
    \caption{Spatial distribution of intralayer exciton wavefunctions of WSe$_{2}$ for a fixed hole with relaxation reduced to one quarter in parallely aligned WS$_{2}$/WSe$_{2}$. We label the three peaks as I$^\prime$, II$^\prime$, and II$^{"}$, which are separated by 133 meV. Peaks I$^\prime$ and II$^\prime$ remain largely detrapped, while peak III$^\prime$ localizes on the AA stacking region. None of the peaks exhibit charge-transfer character. These calculations do not account for additional screening from hBN encapsulation.}
\end{figure}
\clearpage
\newpage

\section{F. Isotropy in optical conductivity of intralayer moiré excitons}
\begin{figure}[ht!]
    \centering
    \includegraphics[scale=0.6]{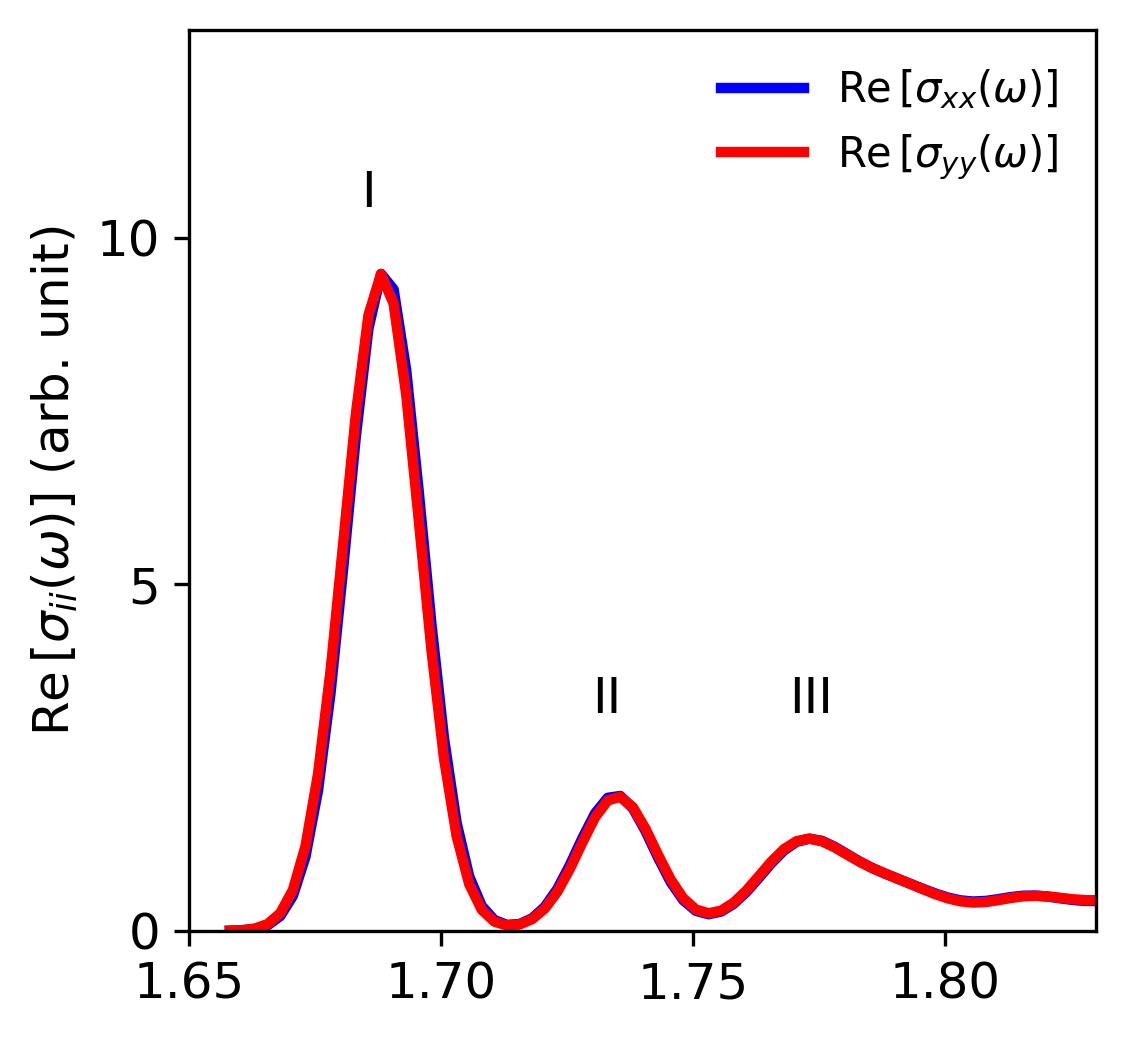}
    \caption{The optical conductivity is isotropic for moir\'e intralayer excitons in a parallel-aligned WS$_2$/WSe$_2$ heterobilayer.}
\end{figure}

\clearpage
\newpage

\bibliography{origin}

\end{document}